# Overcoming the language barrier in mobile user interface design – A case study on a mobile health app

Jason Ross
School of Information Technology and Mathematical Sciences
University of South Australia
Adelaide, Australia
Email: jason-ross1@hotmail.com

Jing Gao
School of Information Technology and Mathematical Sciences
University of South Australia
Adelaide, Australia
Email: jing.gao@unisa.edu.au

## Abstract

This research report proposes a structured solution to address the need for awareness of cultural and language in user design. It will include evaluated research on established methods that already exist. Discussed ideas about how to address this situation include: what others have found to take into consideration when using design principles to develop an interface, detailed troubles and critical issues that have been previously identified and also ways that have been found already to overcome such issues. This will also involve designing a prototype application catering to resolving these issues. Overcoming the language barrier plays an important role in the process of implementing a user design interface that will satisfy users. This issue must be researched and examined to identify the issues and concerns associated in order to provide a solution in an ethical manner.

**Keywords**

**language barrier, mobile user interface, interface design, mobile interface, interface design**.

## 1   Introduction

Businesses have increasing potential to gain advertising, awareness and online transaction due to the ubiquitous use of mobile phone applications in our contemporary lifestyle. Usage of smart phones has increased rapidly all around the globe (Do, Blom et al. 2011) .Falaki, Mahajan et al. (2010) state "By 2011, smart phone sales are projected to surpass desktop PCs". A study was done examining how many applications had been downloaded over the past two years, results displayed 300 million apps downloaded in 2009 to five billion in 2010(Boulos, Wheeler et al. 2011).

The user interface is one of the most important elements to consider when designing applications for businesses because it connects the user to the service they require(Faghih, Azadehfar et al. 2014). The means of obtaining a required service needs to be presented in a user-friendly way; otherwise the user would have insufficient information to navigate the application. The term 'human computer interaction'(HCI) is used to study these interactions (Carroll 2009). From this study in HCI a number of standard guidelines have been developed to guide the design process.

In health services there are some common issues that arise when services are adapted for mobile devices. To address a wide variety of audiences from different social and cultural backgrounds, many possibilities must be considered. Information that is displayed to respective cultures must be examined to ensure that it does not offend anyone but still provides the needed information. Keeping information short and concise is necessary so that it can be navigated to without vast amounts of reading. This can be difficult especially when health service applications are designed to provide information. The designer must pay specific attention to balance provision of information while concisely imparting it.

## 2   Research Question and Design

It is crucial to examine what kind of ideas and rules should be applied to making a mobile website or mobile user interface for specific services, especially in the health domain. This research will examine what needs to be achieved in order to do so. Therefore the research question is as follows: what rules should be applied when designing a mobile application interface for a culturally diverse clientele?

This research was conducted as a case study in a public hospital where a mobile application was designed and implemented to provide post-natal breast feeding advice to all patients who have diversified cultural background. Through interactive design and pre and post development surveys, answers were collected to address the research question.

# 3 Literature Review

## 3.1 Overview of Mobile Systems and User Interface Design

The popularity of smart mobile devices is increasing and one major contributing factor to this is the connectivity to businesses and organisations. This is evident in the health industry; from a survey conducted in 2012 31% of smart phone owners use their phones to look up medical or health information online. This is an increase of 17% compared to users in September 2010 (Fox and Duggan 2012). Factors contributing to smart phone usage might be different for all cultures. Sarker and Wells (2003) state "Various factors such as demographics, technology-related skills, and culture were identified as important determinants influencing the implementation and acceptance of wireless handheld phones."

Mobile application usage appeals to a wide audience of consumers, particularly in the health industry. This is because it is now possible for a user to store and refer to information, save data that is critical, and manipulate this data to achieve complex calculations. It is now easier to access content from the internet with the ability to display it in video and audio media (Franko and Tirrell 2012). All these features have produced positive outcomes within the health industry. Van Velsen, Beaujean et al. (2013) state: "For medical professionals, the use of mobile technology has been found to be beneficial, as it allows them to make decisions more rapidly and with a lower error rate, and to increase the quality of data management and data accessibility. For patients, mobile technology improves patient education, self-management of chronic diseases and it greatly enhances the possibilities for remote monitoring of patients." Although many positive aspects may be found by mobile applications in health, there are problems that must be recognised consisting of reading from a small screen, slow download speed, and troublesome input mechanisms (van Velsen, Beaujean et al. 2013).

The design and look of a website or mobile application is of high importance when implementing a specific service. Content and design are the most important aspects in attracting users to websites. If a website is aesthetically pleasing a user will be more inclined to visit it again (Venkatesh and Ramesh 2006). If a website does not appeal to its audience and the user does not enjoy the experience, the service will less likely be revisited. From this Seyedi, Taherzadeh et al. (2011) suggest that"identification of qualitative factors effective on website quality is vital for survival and improvement".

It is vital to consider which aspects of the layout design affect the cognitive approach users take while interacting with the interface. (Dillon 2003) lists perception, memory, attention, categorisation, decision-making, comprehension and related processes as key factors to understand when taking into account how to design interfaces. Dillon (2003) also states "for interfaces we need to consider if users will perceive actions and behaviours, how they will interpret them, what demands our designs place on their attention, what knowledge they will need to interpret and respond in a manner acceptable to the software."

Interface designers have worked with industry organisations for many years in order to refine the process so that users can work with technology to access the services they require. Research is still integral in the foundation for the layout of specific business interfaces as the market is constantly changing and the consumer is perennially evolving. Lee and Kozar (2006) state "During the past decades, companies made large investments in the implementation of information systems with the expectation of productivity gains, competitiveness enhancement, and the reduction of market, administrative and operational costs. However such claims have not been validated by empirical data. Therefore, researchers have made efforts to propose a better way of evaluating information systems."

A number of standard guidelines have been developed to guide the design process of proper human computer interaction. These give designers a medium to follow when bearing in mind user interface design. One set of guidelines is titled the "7C's". This involves critically thinking about the Context, Communication, Connection, Commerce, Content, Community and Customisation in detail to prepare for designing an application (Lee and Benbasat 2003). When considering the UI it is important to note how the style of content is laid out. The usability of a proposed interface is impacted by whether good consistency is implemented (Kim and Foley 1993). Keeping information straight to the point and to a minimum is a good way to keep the user interested. On this Nielsen (1999) makes an important point in saying "We need ways of summarising large collections of information objects". Another instance of a design guideline is a design tool called "DON" This design tool involves designers identifying high level preferences for their interface design such as the margin width of content. From this the different stages in their design phase will work around these high level preferences to ensure that they are accommodating for all aspects of the design phase (Kim and Foley 1993).

For the user, intuition is important for the interface, if a user is able to use their perception to successfully learn and navigate with ease, a positive experience will be achieved (Lumsden and Global 2008). The user needs the ability to navigate to the content they desire with ease and without getting lost. Using icons and friendly messages is a good strategy in user interfaces. Visual mnemonic icons are a standard that are useful in a range of situations, as there are some icons that people from all around the world will understand. One such instance of this consists of images of a floppy disk for saving and a folder for opening. Egger (2001) believes that we should "Let the customer be in control: Support the browsing behaviours of both novice and expert users, inform customers about the procedures required to transact: e.g. overview of steps, provide clear feedback to user actions: allow for easy error management."

Standards for mobile applications are constantly changing and demand frequent analysis of overall design and usability. Responsiveness is a growing factor contributing to usability; users expect applications to work in a nimble and concise fashion across a range of different devices with varying screen sizes. As there is a large amount of applications being released every day it has caused potential dangers in the health industry amongst those seeking medical advice. There is an overload of medical applications available for download on the different app stores which may cause users confusion as to which ones are reliable sources. There have been initiatives to counter measure the potential of false information acquired from these applications. These initiatives analyse medical applications in order to clarify whether or not the quality of content is reliable (van Velsen, Beaujean et al. 2013).

## 3.2  Rectifying Cultural Design for Mobile User Interface

It is difficult to find a medium for which people from all cultural diversities with different languages can access and use with the same features. The categories and sub categories of online navigation structures are important to consider because they may strongly influence the online behaviour of cultural differences (Venkatesh, Ramesh et al. 2003). On the importance of usability between cultures Kim, Kim et al. (2003) state "Cultural usability is just as important as any other usability issues. Outcomes of cultural studies can provide basic principles and understandings to designing cultural usability."

Known cultural mobile interface issues begin with technical systems using functional slang and having it brought across a language plain.Vandevelde and Van Dierdonck (2003) state "A tower of Babel syndrome may occur. The exchanged information may be interpreted incorrectly or misunderstood. Hence, the value of the information is deemed to be low or non-existent in the eyes of the message receiver. Furthermore, subtle language differences may imply vastly different solutions that may determine whether a project is successful or not."

Cultural differences are an impending problem within interface development and become an issue for marketing the product. An importance for cross cultural research is imminent in allowing a comprehensible understanding to make a better solution for marketing the application (Eune and Lee 2009).Creating a design that is accessible for a variety of audiences is a big issue in user interfaces. When undertaking this task, Dunlop and Brewster (2002) suggest taking into consideration that one is "designing for a widespread population: Users will not normally have any formal training in their technologies and consider them as devices to be used rather than computers to be maintained".

Research has been done as to what kind of medium is currently standing within interface design, in order to overcome the language barrier. Xie, Rau et al. (2009) state "Using succinct information along with visual representations ensures that all cultural backgrounds have an understanding of information they are able to acquire from a mobile application". Simplistic style navigations and avoiding ambiguous topics and links are a must for ease of access. One piece of research by Choi, Lee et al. (2005) comparing Japanese, Korean and Finnish users displayed such results. "The Finnish user cited logical ordering of the menu list as a feature that helped him achieve his goal in a short period of time.  His reference to a logical ordering of the menu corresponds to the characteristics of the monochromic group, namely, high sensitivity to the delay time. "

### 3.2.1 Icons

Icons become a larger issue when considering an application designed for multiple cultures. The recognition of icons between different cultures has be analysed; these studies found that there are some difficulties in recognising certain icons in different situations. The study found that Korean subjects recognised more standard icons more than the American subjects, whereas the American subjects recognised the abstract icons a lot more (Kim and Lee 2005). Jagne (2004) researched and found that "Studies into the impact of new technologies, show that foreign users will show resistance to and reject products with Western metaphors in favour of products localised according to their cultural customs, idioms etc."

### 3.2.2 Layout

If the intended users for an application contain many different cultural backgrounds the layout needs to be kept simple with clear indication of what can be done on the screen and what the navigational features do in order for it to satisfy everyone. Some cultures read information from right to left and view content in a different manner than others. In order to assure that a wide range of users may be able to use the application it is vital for the layout to follow a specific pattern that is recognisable amongst other applications. Information displayed in a vertical fashion with simple button navigation is important.

### 3.2.3 Navigation

If a standard medium of navigational control is implemented via maintained technical communication across multiple-language interfaces, it will be strongly sufficient for all audiences. Xinyuan (2005) advises use of "Simplicity, with clear metaphors, limited choices, and restricted amounts of data". On this Hamilton (2000) adds that "Mismatches that contradict the users' expectations of metaphors, tends to create cognitive conflict, which in turns leads to decreased performance levels". This problem needs to be addressed by cultural and language differences researched from experts amongst such linguists. This issue also needs to address abstract concepts and information within the interface that will likely cause confusion amongst different people.

An editable structure with different dimensions to account for multicultural design is key. In regards to mobile application lifespan, the proposed software needs to be suitable for rapid change to ensure that it will be able to successfully implement another language or change specific elements to satisfy a cultural issue (Badre 2001).

### 3.2.4 Content

Information displayed on an interface needs to be concise and summarised accordingly. If the content is abstract it becomes difficult to translate this information across cultural platforms (Vandevelde and Van Dierdonck 2003). If the content on a website is analysed properly, then this can be prevented. Marcus and Gould (2000) list the metaphors, mental model, navigation, interaction, or appearance to be factors to analyse that could have the possibility of offending or confusing a user from a different cultural background.

A potential issue for language involves the simplicity in information that is being communicated through an application. If a design such as the prototype instance is implemented there is a restriction of how intricate or in depth the detail of the information can go. If the health application requires very concise information it will not be implemented due to potential reading disabilities of the user. This could greatly hinder users from many different cultures. It is very important for images and videos to be implemented along with basic text to overcome this issue.

### 3.2.5 Images

Critical thinking needs to be included when considering a cultural interface design; it cannot just cover the basic elements. Jagne and Smith-Atakan (2006) state "In order to gain a market advantage, companies thought it would be enough to just translate language, currency, date and time formats etc." These factors are important but only the top level of things to consider. On using a cultures flag to associate with recognising their specific culture to attribute with a factor of the website Barber and Badre (1998) found "The flag serves as a symbol of immediate national, even global, recognition, helping the user to quickly identify the locale and origin of the site, which is particularly helpful when the site is in a language foreign to the user. The flag is also used to denote alternative language choices, which impacts usability in that the user may identify and choose an alternate language much more quickly and efficiently, as opposed to when the choices are textual". Imagery may also be offensive to different cultures dependent on the content displayed. If health application imagery consists of diagrams visually representing the human body it may be perceived in different ways.

### 3.2.6 Colour

Colour is an important aspect of cultural design, and which colours are appealing to a specific culture needs to be addressed. Plass (1998) recommends to: "Follow perceptual guidelines for good colour usage. Examples: use warm colours for advancing elements and cool colours for receding elements; avoid requiring users to recall in short-term memory more than 5±2 different coded colours." The design for cross cultural applications needs to be investigated despite the difficult nature of appealing to all audiences (Walton, Vukovic et al. 2002). One such study by Lee, Choi et al. (2008) whereupon ten different cultural backgrounds used the same user interface found "The ten cultural dimensions were to be affected substantially according to the countries where users resided and the devices which they were using."

This kind of information assists with providing an understanding of how current operations work although evidently cannot be incorporated into every scenario. Various cultural differences have to be taken into consideration as to not offend any user. For example the colour green could be used for a health based application although it may offend people from France as it is associated with criminality, whereas in Egypt and Middle-Eastern countries, green is associated with positive influence (Barber and Badre 1998).

## 4 Discussion and Findings

A case study was produced in association to the information found regarding health care interfaces and overcoming cultural barriers. This information assisted in creating a cultural design scenario which was analysed on how to overcome these issues and then implement them into an application for a health service. From this scenario a mobile application prototype was designed for providing new mothers with information on postnatal care.

Firstly the participants in this study including hospital staff and postnatal patients provided some information on which systems to design for and their opinions on what popular colours and themes would seem appropriate to them.

This information concluded that green or blue is a better choice for a friendly tone when considering a health application. Staying away from the colour red is a good choice as it is often associated with danger fear, or blood in the instance of a hospital. These friendlier colours assisted with the project application in allowing it to be more user-friendly.

It was discovered that a majority of postnatal patients use iPads to connect to businesses and software applications, from this result the design must be fully functional on various iPad devices in particular and aligning with Apple's application development policies and guidelines.

A summary of mobile design consideration and solutions is provided below:

| Issue | Definition | Potential Resolutions | Final Solution |
|---|---|---|---|
| **Representing languages with flags not suited for every language** | Persian users may associate themselves with one of many flags, if the user associates themselves with a flag that is not shown for Persians, they might be offended | Place all flags associated with one language<br>No flags at all<br>Add flags with a "hello" message in their respective language | Removed flags and just have their language followed by "hello" |
| **Colour palette for application** | Required to use a friendly colour basis, examine what would work well for all languages | Keeping away from "hot" colours or intimidating colours, for example: red yellow or orange<br>Research across multi | Working with a green or blue is a good choice to associate with friendliness within a health care application. Using different blues |

| | | language interfaces for a standard ground | was implemented into this design. |
|---|---|---|---|
| **Which device is most accessible to postnatal patients** | A means for the device use most is necessary to test it works with the application sufficiently | Acquire data from postnatal patients | Data was acquired resulting in iPads being the preferred method of using applications. |
| **Background image implemented into application** | Background images were evident in the prototype designs | Survey hospital staff and professionals for opinions | No background image was final decision- enables viewing information possible without distractions |
| **Icons for certain navigational elements** | To satisfy all cultural differences, standard icons that can be recognised need to be implemented. | Research what kind of standards are already used in current software applications | These proposed standards were implemented into our own design. |

*Table 1: Mobile UI design issues and possible solutions.*

The prototype application was designed for new mothers during postnatal care. Information from a hospital was provided to be placed into sections of the application as content that the user can view. The application was designed to be intuitive to use with simple navigation to pages containing short sentences and graphical imagery involving postnatal care.

Once downloaded, on start-up it the application prompts the user for their language. It will be easy for the user to recognise their language from the options provided which are displayed in their respective languages. Secondly, the application loads up and presents the information dependant on what language was selected. The list of various topics is displayed for navigation and a user can touch one to view the information for each of them.

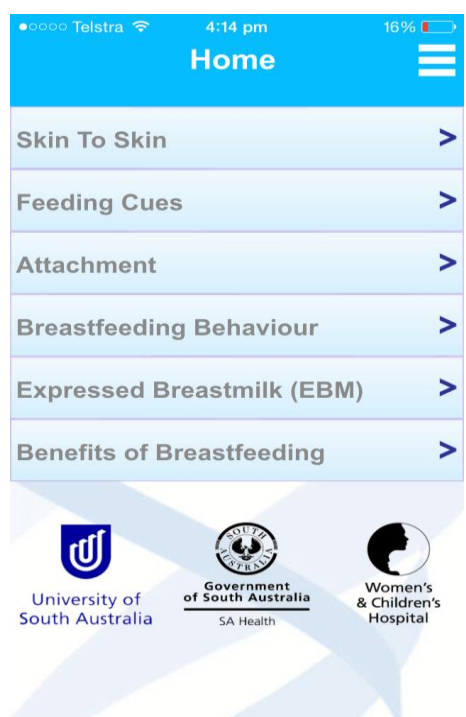

*Figure 1. Mobile Application main page*

If a user wants to change their language they can click on the settings icon in the top right hand corner. This settings image is a worldwide standard and is commonly associated with settings on a majority of modern popular applications or websites. In the settings page there are switches to toggle larger font, internet downloading and language selection. Since this application will be primarily accessed offline the information will be stored in the cache of the phone. If there is an updated version of the content the user will be prompted to download the new content on start-up.

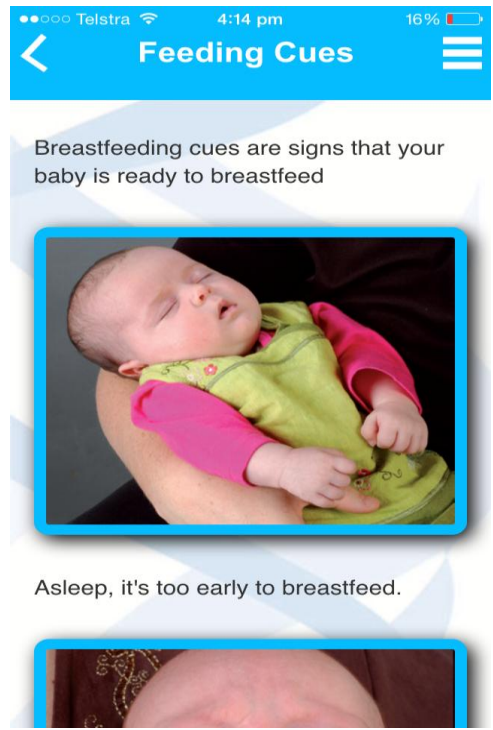

*Figure 2. Mobile Application content and imagery*

This application was specifically for users of 8 different languages and cultures within health care consisting of: Mandarin, Dari, Hindi, Persian, Aboriginal specifically Pitjantjatjara, Australian, Arabic and Vietnamese. These factors were widely considered when the prototype was designed to ensure that the application was usable by all audiences. Intuitiveness being one of the most important factors was heavily implemented through buttons with specific titles to ensure the user that they knew where they were navigating to. The layout was kept consistent with the overall theme displaying in all pages. The navigation was smooth and delivered effectively through animations that kept the application robust and nimble in the way it was presenting its data.

The experiment did not assume more than basic reading skills amongst the subjects. Research in the design for usability was accomplished to an optimal standard as followed by incorporating all the guidelines read from specific papers mentioned in the literature review. The standard of the application was produced incorporating everything involved within this study. This included information in the literature and case study with the questionnaires. In result the application was designed with intuitiveness, imagery associated with small paragraphs of text and a navigational structure that lets the user view exactly what they intended. Graphical icons were implemented portraying navigation options and a simple page layout without too much information on it kept the user focused on what was on the screen to minimise a loss of interest.

## 5   Conclusion

This research found common issues that arise across the health industry plain when overcoming language barriers for mobile interfaces. As a means to overcome these issues, it is necessary for the design phase to include all possible cultural differences to implement navigational structures and general design principals to satisfy clients on a global scale.

Mobile applications have many factors that must be assessed during the design phase. Factors such as users, simplicity, layout, colours, text and accessibility are all standard considerations to take, if designing for multiple cultures they must also be assessed from this angle. In order to satisfy these requirements across multi-cultural applications, more research must be done in order to analyse the specific cultures that will be using these applications. Once an understanding of all standard requirements of these cultures has been gathered, then a culturally diverse application may be designed.

# 6   References

Badre, A. (2001). "The effects of cross cultural interface design orientation on World Wide Web user performance."

Barber, W. and A. Badre (1998). Culturability: The merging of culture and usability. Proceedings of the 4th Conference on Human Factors and the Web.

Boulos, M. N., S. Wheeler, et al. (2011). "How smartphones are changing the face of mobile and participatory healthcare: an overview, with example from eCAALYX." Biomedical engineering online **10**(1): 24.

Carroll, J. M. (2009). "Human–Computer Interaction." Encyclopedia of Cognitive Science.

Choi, B., I. Lee, et al. (2005). A qualitative cross-national study of cultural influences on mobile data service design. Proceedings of the SIGCHI conference on Human factors in computing systems, ACM.

Dillon, A. (2003). "User interface design." Encyclopedia of Cognitive Science.

Do, T. M. T., J. Blom, et al. (2011). Smartphone usage in the wild: a large-scale analysis of applications and context. Proceedings of the 13th international conference on multimodal interfaces, ACM.

Dunlop, M. and S. Brewster (2002). "The challenge of mobile devices for human computer interaction." Personal and Ubiquitous computing **6**(4): 235-236.

Egger, F. N. (2001). Affective design of e-commerce user interfaces: How to maximise perceived trustworthiness. Proc. Intl. Conf. Affective Human Factors Design.

Eune, J. and K. P. Lee (2009). "Analysis on Intercultural Differences through User Experiences of Mobile Phone for globalization." Proceedings of International Association of Societies of Design Research (Coex, Seoul, Korea.

Faghih, B., D. Azadehfar, et al. (2014). "User Interface Design for E-Learning Software." arXiv preprint arXiv:1401.6365.

Falaki, H., R. Mahajan, et al. (2010). Diversity in smartphone usage. Proceedings of the 8th international conference on Mobile systems, applications, and services, ACM.

Fox, S. and M. Duggan (2012). Mobile health 2012, Pew Internet & American Life Project Washington, DC.

Franko, O. I. and T. F. Tirrell (2012). "Smartphone app use among medical providers in ACGME training programs." Journal of medical systems **36**(5): 3135-3139.

Hamilton, A. (2000). "Metaphor in theory and practice: the influence of metaphors on expectations." ACM Journal of Computer Documentation (JCD) **24**(4): 237-253.

Jagne, J. (2004). "Integrating cultural and social factors of the shopping metaphor, in the context of indigenous users, into ecommerce interface design." Interaction Design Centre, Middlesex University.

Jagne, J. and A. S. G. Smith-Atakan (2006). "Cross-cultural interface design strategy." Universal Access in the Information Society **5**(3): 299-305.

Kim, J. H. and K. P. Lee (2005). Cultural difference and mobile phone interface design: Icon recognition according to level of abstraction. Proceedings of the 7th international conference on Human computer interaction with mobile devices & services, ACM.

Kim, S., M. Kim, et al. (2003). "Cultural issues in handheld usability: Are cultural models effective for interpreting unique use patterns of Korean mobile phone users." Proceeding of UPA (Usability Professionals' Association).

Kim, W. C. and J. D. Foley (1993). Providing high-level control and expert assistance in the user interface presentation design. Proceedings of the INTERACT'93 and CHI'93 Conference on Human Factors in Computing Systems, ACM.

Lee, I., G. W. Choi, et al. (2008). Cultural dimensions for user experience: cross-country and cross-product analysis of users' cultural characteristics. Proceedings of the 22nd British HCI Group Annual Conference on People and Computers: Culture, Creativity, Interaction-Volume 1, British Computer Society.

Lee, Y. and K. A. Kozar (2006). "Investigating the effect of website quality on e-business success: an analytic hierarchy process (AHP) approach." Decision support systems **42**(3): 1383-1401.

Lee, Y. E. and I. Benbasat (2003). "Interface design for mobile commerce." Communications of the ACM **46**(12): 48-52.

Lumsden, J. and I. Global (2008). Handbook of research on user interface design and evaluation for mobile technology, Information Science Reference.

Marcus, A. and E. W. Gould (2000). "Crosscurrents: cultural dimensions and global Web user-interface design." interactions **7**(4): 32-46.

Nielsen, J. (1999). "User interface directions for the web." Communications of the ACM **42**(1): 65-72.

Plass, J. L. (1998). "Design and evaluation of the user interface of foreign language multimedia software: A cognitive approach." Language Learning & Technology **2**(1): 35-45.

Sarker, S. and J. D. Wells (2003). "Understanding mobile handheld device use and adoption." Communications of the ACM **46**(12): 35-40.

van Velsen, L., D. J. Beaujean, et al. (2013). "Why mobile health app overload drives us crazy, and how to restore the sanity." BMC medical informatics and decision making **13**(1): 23.

Vandevelde, A. and R. Van Dierdonck (2003). "Managing the design-manufacturing interface." International Journal of Operations & Production Management **23**(11): 1326-1348.

Venkatesh, V. and V. Ramesh (2006). "Web and wireless site usability: understanding differences and modeling use." Mis Quarterly: 181-206.

Venkatesh, V., V. Ramesh, et al. (2003). "Understanding usability in mobile commerce." Communications of the ACM **46**(12): 53-56.

Walton, M., V. Vukovic, et al. (2002). 'Visual literacy'as challenge to the internationalisation of interfaces: a study of South African student web users. CHI'02 extended abstracts on Human factors in computing systems, ACM.

Xie, A., P.-L. P. Rau, et al. (2009). "Cross-cultural influence on communication effectiveness and user interface design." International Journal of Intercultural Relations **33**(1): 11-20.

Xinyuan, C. (2005). Culture-based user interface design. IADIS AC.

## Acknowledgements

This document was adapted from the Instructions for Authors from ICIS2007 (which in turn was adapted from the AMCIS templates), PACIS 2007, ACIS 2011, ACIS 2010, ACIS 2008, ACIS2007, ACIS 2006, and the ACIS 2005 Instructions, which were an extension of the ACIS 2004 instructions, much of which was adapted from the ACIS 2003 and ACIS 2002 Instructions, which were based on the ACIS'98 Instructions (which was adopted from ACIS'97 Instructions). These in turn were adapted from an "Instructions for Authors" written by Roger Clarke. The new format, use of the Creative Commons license and support for DOIs was added by John Lamp in 2015.

## Copyright